\documentclass[preprint,showpacs,preprintnumbers,amsmath,amssymb]{revtex4}

\usepackage{graphicx}
\usepackage{dcolumn}
\usepackage{bm}

\begin{document}

\title{Electric breakdown in ultra-thin MgO tunnel barrier junctions for spin-transfer torque switching}

\author{M.\ Sch\"afers}
 \email{mschaef@physik.uni-bielefeld.de}
 \homepage{http://www.spinelectronics.de}
\author{V.\ Drewello}
\author{G.\ Reiss}
\author{A.\ Thomas}

\affiliation{Bielefeld University, Department of Physics, Thin Films and Physics of Nanostructures, 33501 Bielefeld, Germany}

\author{K.\ Thiel}
\affiliation{Fraunhofer Institut f\"ur Fertigungstechnik und Angewandte Materialforschung, Wiener Str.\ 12, 28359 Bremen, Germany}

\author{G.\ Eilers}
\author{M.\ M\"unzenberg}
\author{H.\ Schuhmann}
\author{M.\ Seibt}

\affiliation{I.\ \& IV.\ Physikalisches Institut and Sonderforschungsbereich 602, Friedrich-Hund-Platz 1, Georg-August-Universit\"at G\"ottingen, 37077 G\"ottingen, Germany}


\date{\today}

\begin{abstract}
Magnetic tunnel junctions for spin-transfer torque switching were prepared to investigate the dielectric breakdown. The breakdown occurs typically at voltages not much higher than the switching voltages, a bottleneck for the implementation of spin-transfer torque Magnetic Random Access Memory. Intact and broken tunnel junctions are characterized by transport measurements and then prepared for transmission electron microscopy and energy dispersive x-ray spectrometry by focussed ion beam. The comparison to our previous model of the electric breakdown for thicker MgO tunnel barriers reveals significant differences arising from the high current densities. 
\end{abstract}

%
\maketitle
The switching of magnetic tunnel junctions (MTJs) by spin-transfer torque has gained high interest in the last years \cite{huai2004apl84p3118,fuchs2004apl85p1205,hayakawa2005jjap44pL1267}. From a technological point of view this effect can be used for compact microwave oscillators \cite{Deac2008nphys4p4763} or a new type of magnetic random access memory (MRAM) \cite{hosomi2005iedm2005p459,katine2008jmmm320p1217}, which does not need special writing lines above the MTJ cells. This spin-transfer torque (STT) MRAM can be integrated at much higher densities which is essential for applications.

As high current densities are required to switch MTJs with the STT method, MTJs with a thin barrier and, therefore, low area resistance are used. Still a relatively high voltage is applied to the junction to achieve the switching currents. These are in the range of a few hundred mV compared to only a few mV needed for normal (read) operation  \cite{huai2004apl84p3118}. 
The application of higher voltages is limited, as at some point an electrical breakdown of the barrier is observed \cite{oepts1998apl73p2363}. Furthermore, this breakdown happens at lower voltages for thinner barriers \cite{khan2008jap103p4455}. 

The breakdown voltage should be much higher than the switching voltage in order to optimize the stability of the junctions. The breakdown effect has been widely investigated by electrical transport measurements \cite{khan2008jap103p4455, oepts2001jap89p8038, oliver2002jap91p4348, schmalhorst2001jap89p586}. While this makes an optimization with large amounts of samples more practicable it provides little information about the microscopic processes during or after a breakdown. 

In this paper we present transmission electron microscopy analysis of MTJs which were stressed through a dielectric breakdown. They show a very different breakdown behavior compared to thicker barriers \cite{thomas2008apl93p152508}.

The breakdown investigated in this study is commonly called 'hard breakdown'. A sudden increase in the current (during constant voltage stress in this case) indicates this type of breakdown.

The MTJs are sputter deposited in a Singulus \textsc{ndt timaris ii} cluster tool. The layer stack is 5 (nm) Ta/90 Cu-N/5 Ta/20 Pt-Mn/2.2 Co$_{70}$Fe$_{30}$/0.8 Ru/2 Co$_{60}$Fe$_{20}$B$_{20}$/1.1 MgO/1.5 Co$_{60}$Fe$_{20}$B$_{20}$/10 Ta/30 Cu-N/7 Ru. The complete stack is annealed for 90 minutes at 360$^\circ$C in a magnetic field of 1\,T. 

Elliptic junctions sized 360\,nm $\times$ 150\,nm are patterned into a resist layer by conventional electron beam lithography. Argon ion beam etching transfers the patterns into the layer stack. The complete sample is covered with a thick layer of tantalum oxide to electrically insulate the separated pillars from each other. The top of the pillars are opened with a lift-off process. 
A gold layer is deposited as upper lead material. Contact pads are created by a second exposure and etching process.

The samples for high-resolution electron microscopy (HRTEM) investigation have been prepared by focused ion beam (FIB) with a FEI Nova Nanolab 600 instrument. Damage to the device caused by the Ga$^+$ ions during milling was prevented by a 0.5\,$\mu$m Pt coating on top of the lamella. A last low energy etching step at 5\,kV under an incidence angle of 7$^\circ$ was performed to reduce the amorphous surface layer resulting from the cut at 30\,kV Ga$^+$ beam. The final thickness of the lamella is in the range of 10-20\,nm. TEM work was done using a Philips CM200-FEG UT operated at an accelerating voltage of 200kV. The microscope has a point resolution of 0.19\,nm and an information limit of 0.11\,nm. Energy-dispersive x-ray spectrometry (EDX) was performed using a Si:Li detector (Link ISIS).

\begin{figure}
\includegraphics{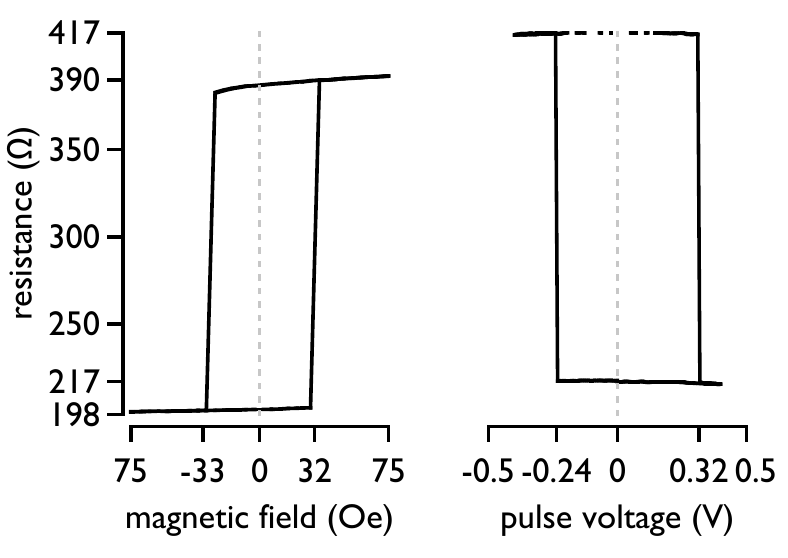}
\caption{\label{fig:stt_intact}Magnetic minor loop and resistance vs.\ voltage plot of the magnetic tunnel junction shown in the left images of Fig.~\ref{fig:tem} and Fig.~\ref{fig:edx}.}
\end{figure}

About 200 magnetic tunnel junctions were prepared by e-beam lithography and examined by conventional 2-terminal transport investigations to determine the characteristics of the junctions. Figure~\ref{fig:stt_intact} (left) shows the magnetic minor loop of one of the junctions. A TMR ratio of 97\% was reached. This is a typical value for low resistive tunnel barriers for STT switching \cite{kubota2005jjap44pL1237,serrano-guisan2008prl101p087201}. 

The corresponding STT loop is shown on the right hand side of Figure~\ref{fig:stt_intact}. The TMR ratio is now 92\% due to slightly changed resistances in the parallel and anti-parallel states. This is caused by diffusion processes and explained for these junctions in detail by Krzysteczko et al.\ \cite{krzyszteczko2009jmmm321p144} The switching voltages of $-0.24$\,V and $0.32$\,V correspond to switching current densities of  $-2.7\times10^{6} \rm{A/cm^{2}}$ and $3.8\times10^{6} \rm{A/cm^{2}}$, respectively. 

\begin{figure}
\includegraphics{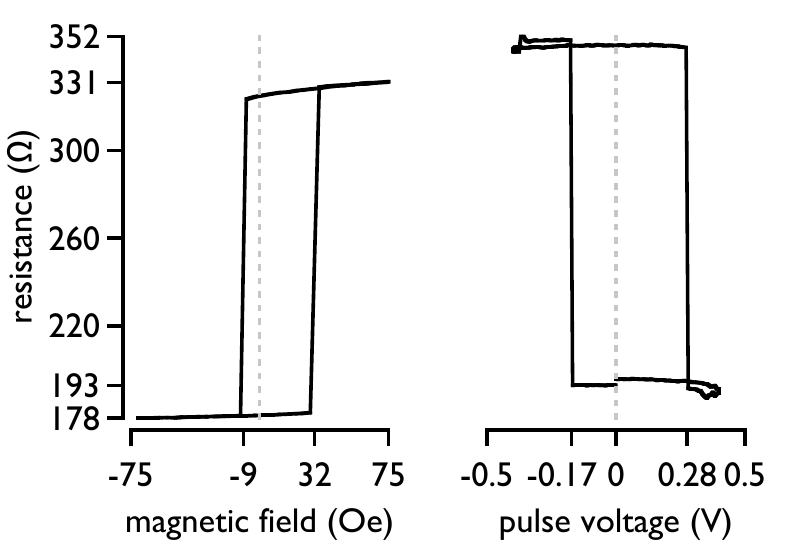}
\caption{\label{fig:stt_broken}Magnetic minor loop and resistance vs.\ voltage plot of the MTJ shown in the right images of Fig.~\ref{fig:tem} and Fig.~\ref{fig:edx} {\em before} it was stressed through the dielectric breakdown.}
\end{figure}

Figure~\ref{fig:stt_broken} depicts the magnetic minor and STT loops of another magnetic tunnel junction. Here, the TMR ratio is 86\% for the magnetic loop and 82\% in the STT case. The current densities are $-2.2\times10^{6} \rm{A/cm^{2}}$ switching from parallel to anti-parallel and $3.7\times10^{6} \rm{A/cm^{2}}$ vice versa. A few junctions were stressed through a dielectric breakdown after the measurements including the junction investigated in Figure~\ref{fig:stt_broken}.

\begin{figure*}
\includegraphics[width=\linewidth]{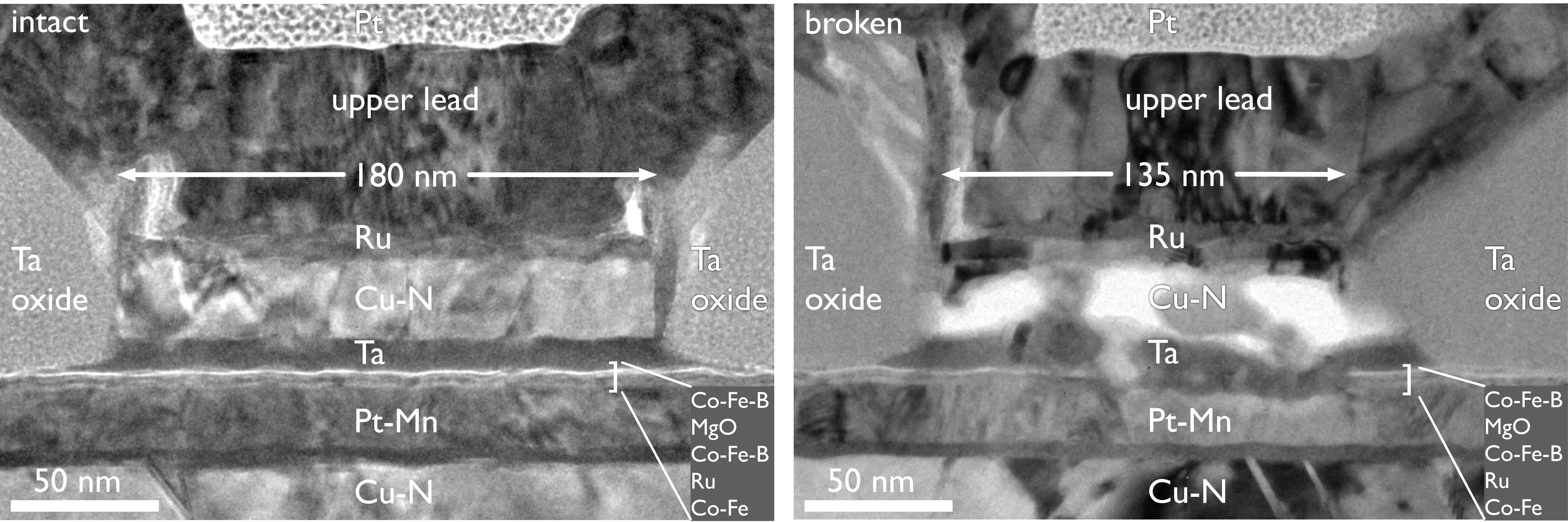}
\caption{\label{fig:tem}Transmission electron microscopy image of an intact (left) and a broken (right) magnetic tunnel junction. The Ta oxide and the upper lead are necessary for the definition of the junction by the lithography processes.}
\end{figure*}

The left hand side of Figure \ref{fig:tem} shows a TEM image of the junction characterized in Figure \ref{fig:stt_intact}. The complete layer stack excluding the lower Ta layer is visible. The Ta oxide on the left and right hand side of the image and the upper lead are necessary for the preparation of the junction by lithography. The junction width as seen in the TEM data depends on the position where the ellipsis was cut by the focussed ion beam. The thin bright white line just underneath the Ta oxide is the MgO tunnel barrier. It extends through the full size of the image. We verified the smoothness of the three dimensional extension of the MgO barrier into the of 10-20\,nm thick lamella by transmission electron tomography. See EPAPS supplementary material at [URL will be inserted by AIP] for a movie of transmission electron tomography reconstructed from TEM images.

The same target preparation was applied to the stressed junction characterized in Figure~\ref{fig:stt_broken}. The TEM image of this junction is depicted on the right hand side of Figure~\ref{fig:tem}. All TEM images and EDX scans are rotated and scaled to reveal the same region of the layer stack for a better comparison.

The most dramatic change is seen in the region of the upper Cu-N. Within the bright regions, all material has been competely removed during breakdown. A closed up look to the tunnel barrier region reveals that the thin white line, the signature of the MgO barrier, can not be identified on a width of about 80\,nm. One concludes that the barrier within that region has completely vanished.

\begin{figure*}[t]
\includegraphics[width=\linewidth]{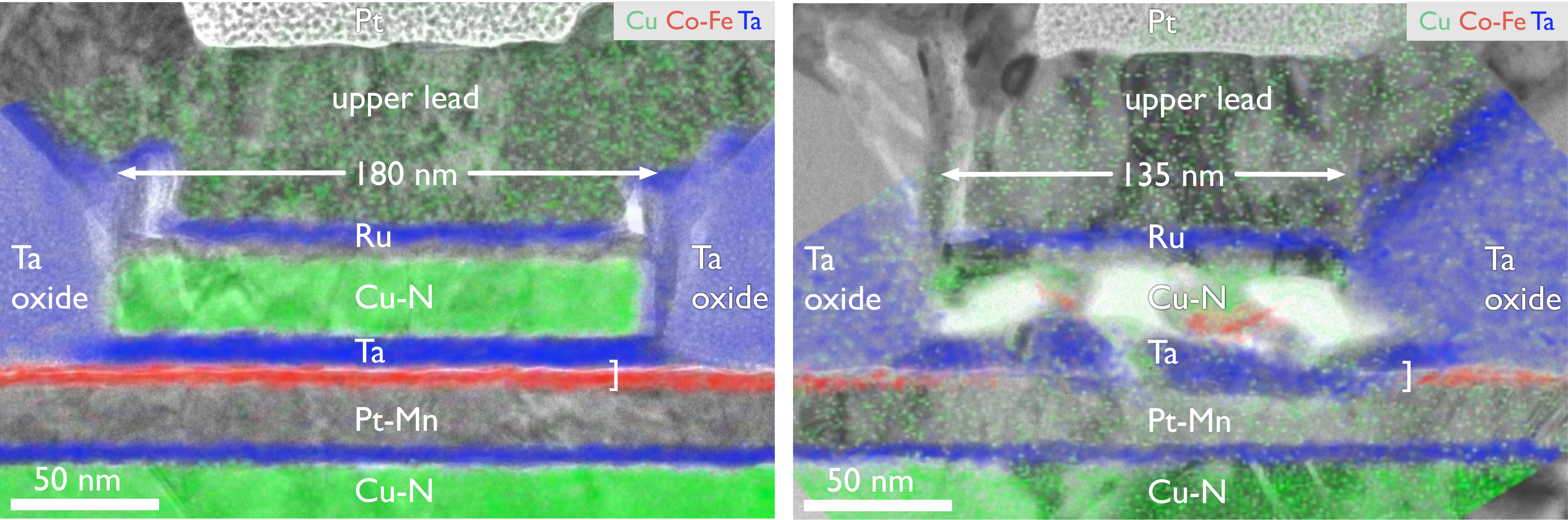}
\caption{\label{fig:edx}EDX scan across the TEM image of an intact (left) and a broken (right) magnetic tunnel junction overlaid to the transmission electron microscopy image of an intact (left) and a broken (right) magnetic tunnel junction. Cu, Co-Fe and Ta distributions are mapped in the image.}
\end{figure*}

To solidify this conclusion drawn from the TEM images, we carried out energy dispersive x-ray (EDX) investigations of the TEM slices to yield element specific images. The left image in Figure~\ref{fig:edx} shows the intact junction. Only Cu, Co-Fe and Ta are mapped here. The lateral resolution of the EDX is limited, therefore, the ferromagnetic electrodes are not separated from the tunnel barrier, but visible as a thick red stripe.

The extent of the damage after breakdown is obvious in the EDX map of the tunnel barrier region and the adjacent layers in the right image of Figure~\ref{fig:edx}. Not only the tunnel barrier but the ferromagnetic electrodes are affected by the high current density of $7\times10^{6} \rm{A/cm^{2}}$ during and $9\times10^{7} \rm{A/cm^{2}}$ after breakdown. Within the missing barrier, the material is completly displaced by the local heating and electrostatic forces ("electron wind") arising from the $\rm{e^-}$-flux from the bottom to the top originating from the high current densities. The electrode material is found 40\,nm above the position of the element (Figure~\ref{fig:edx}, right). We compared the different junction types in table~\ref{tab:comparison} including alumina based junctions \cite{oepts1998apl73p2363,shimazawa2000jap87p5194} and thicker (2.0\,nm) MgO based systems \cite{thomas2008apl93p152508}.

\begin{table}[h]
\begin{tabular}{lcccc}
\hline
barrier material & &AlO$_{\rm x}$ & MgO & MgO \\
barrier thickness & nm & 1.8 & 2.0& 1.1\\
\hline
junction area & $\,\rm\mu m^2$& 10k & 90 & 0.044 \\
area resistance at 10\,mV &$\,\rm\Omega\mu m^2$ &10M & 50k & 8 \\ 
breakdown (bd) voltage &V& 1.5 & 1.6 & 0.4 \\
resistance at bd & $\Omega$ & $\sim 600$ & 330 & 130 \\
resistance after bd & $\Omega$ & $\sim 10$& $\sim 10$& $\sim 10$ \\ 
current density at bd & $\rm{A/cm^{2}}$& 24 & 5k & 7M \\
current density after bd & $\rm{A/cm^{2}}$ & ~$\sim$ 1.5k~ & $~\sim$ 180k~ & ~$\sim$ 90M~ \\
\hline
\end{tabular}
\caption{Comparison of the three systems. The resistance after breakdown is determined by the lithographic leads. The resistance of the junctions should be between 100\,$\Omega$ and 1000\,$\Omega$ to neglect this serial resistance before breakdown.}
\label{tab:comparison}
\end{table}

Even though the mechanisms of the electric breakdown are similar, the consequences arising from the different area resistances are very different. Single pinholes were reported in the case of alumina. We showed in our earlier work that multiple pinholes were found in the case of the thicker MgO \cite{thomas2008apl93p152508} and explained the effect by modeling the local current distribution. Local heating resulted in a crystallization of the electrodes above the pinholes. The pinhole distance was found to be about 50 to 100\,nm. Here the structural confinement does not allow to form multiple pinholes in the element. The numerous pinholes merge and form a large break in the tunnel barrier. Furthermore drastic effects due to the increased current density of $9\times10^{7} \rm{A/cm^{2}}$ (MgO 1.1\,nm MgO) as compared to $1.8\times10^{5} \rm{A/cm^{2}}$ (2.0\,nm MgO) have three possible consequences: The temperature increase is large enough to either heat the element significantly above the Co-Fe-B crystallization point or even local melting. Also, the strong local electrostatic forces at the barrier that act on the TMR stack may lead to electromigration. The $\rm{e^-}$-flux direction from the bottom to the top determines the direction of the material transport. The relative contributions of these processes to the microstructural changes can not be deduced from the present study and need further investigations.

In summary, we presented electric characterizations and TEM images with EDX scans of magnetic tunnel junctions showing current induced magnetization switching. The junctions with ultra-thin tunnel barriers were investigated before and after a dielectric breakdown. We observed a breakdown of the tunnel barrier over several 10\,nm and damage of the adjacent ferromagnetic electrodes. It can be explained by the large current density and lateral confinement in case of the spin-transfer toque devices.

\begin{acknowledgments}
We acknowledge J.\ Schmalhorst for helpful discussions and Singulus Nano Deposition Technology for providing the layer stacks. Support by the Deutsche Forschungsgemeinschaft within the priority program SFB 602 and the research grant \#RE1052/13 is gratefully acknowledged. 
\end{acknowledgments}



\end{document}